\documentclass{aastex63}
\usepackage{graphics,graphicx}
\usepackage{helvet}
\usepackage{hyperref}
\usepackage[utf8]{inputenc}
\usepackage[T1]{fontenc}
\usepackage{soul}
\usepackage{amsmath, amssymb, gensymb}
\usepackage{ulem}
\usepackage{natbib}
\usepackage{microtype}
\usepackage{multirow}
\usepackage{morefloats}
\usepackage{rotating}

\newcommand{\kms}{\mbox{km\,s$^{-1}$}}

\newcommand{\x}{\mbox{$\times$}}
\newcommand{\Msun}{\mbox{M$_{\odot}$}}

\newcommand{\E}[1]{\hbox{$10^{ #1 }$}}

\begin{document}


\title{The Explosion in Orion-KL as Seen by Mosaicking the Magnetic Field with ALMA}
\shorttitle{The Bomb}
\shortauthors{Cort\'es et al.}

\author[0000-0002-3583-780X]{Paulo C. Cortes}
\affiliation{National Radio Astronomy Observatory, 520 Edgemont Road, Charlottesville, VA 22903, USA}
\affiliation{Joint ALMA Observatory, Alonso de C\'ordova 3107, Vitacura, Santiago, Chile}

\author[0000-0002-5714-799X]{Valentin J. M. Le Gouellec}
\affiliation{European Southern Observatory, Alonso de C\'ordova 3107, Vitacura, Santiago, Chile}
\affiliation{Universit\'e Paris-Saclay, CNRS, CEA, Astrophysique, Instrumentation et Mod\'elisation de Paris-Saclay, 91191, Gif-sur-Yvette, France}

\author[0000-0002-8975-7573]{Charles L. H. Hull}
\affiliation{National Astronomical Observatory of Japan, NAOJ Chile, Alonso de C\'ordova 3788, Office 61B, 7630422, Vitacura, Santiago, Chile}
\affiliation{Joint ALMA Observatory, Alonso de C\'ordova 3107, Vitacura, Santiago, Chile}
\affiliation{NAOJ Fellow}

\author[0000-0002-3829-5591]{Josep M. Girart}
\affiliation{Institut de Ci\`encies de l'Espai (ICE-CSIC), Campus UAB, Carrer de Can Magrans S/N, E-08193 Cerdanyola del Vall\`es, Catalonia}
\affiliation{Institut d'Estudis Espacials de Catalunya, E-08030 Barcelona, Catalonia}

\author[0000-0003-3814-4424]{Fabien Louvet}
\affiliation{Universit\'e Paris-Saclay, CNRS, CEA, Astrophysique, Instrumentation et Mod\'elisation de Paris-Saclay, 91191, Gif-sur-Yvette, France}
\affiliation{Universidad de Chile, Camino el Observatorio 1515, Las Condes, Santiago de Chile, Chile}

\author{Edward B. Fomalont}
\affiliation{National Radio Astronomy Observatory, 520 Edgemont Road, Charlottesville, VA 22903, USA}
\affiliation{Joint ALMA Observatory, Alonso de C\'ordova 3107, Vitacura, Santiago, Chile}

\author[0000-0002-5158-0063]{Seiji Kameno}
\affiliation{National Astronomical Observatory of Japan, NAOJ Chile, Alonso de C\'ordova 3788, Office 61B, 7630422, Vitacura, Santiago, Chile}
\affiliation{Joint ALMA Observatory, Alonso de C\'ordova 3107, Vitacura, Santiago, Chile}
\affiliation{The Graduate University for Advanced Studies, SOKENDAI, Osawa 2-21-1, Mitaka, Tokyo 181-8588, Japan}

\author[0000-0002-3296-8134]{George A. Moellenbrock}
\affiliation{NRAO Array Operations Center, P.O. Box O, Socorro, NM, USA}

\author[0000-0003-0292-3645]{Hiroshi Nagai}
\affiliation{National Astronomical Observatory of Japan, 2-21-1 Osawa, Mitaka, Tokyo 181-8588, Japan}
\affiliation{The Graduate University for Advanced Studies, SOKENDAI, Osawa 2-21-1, Mitaka, Tokyo 181-8588, Japan}

\author[0000-0002-6939-0372]{Kouichiro Nakanishi}
\affiliation{National Astronomical Observatory of Japan, 2-21-1 Osawa, Mitaka, Tokyo 181-8588, Japan}
\affiliation{The Graduate University for Advanced Studies, SOKENDAI, Osawa 2-21-1, Mitaka, Tokyo 181-8588, Japan}

\author[200~0000-0003-4314-4947]{Eric Villard}
\affiliation{European Southern Observatory, Alonso de C\'ordova 3107, Vitacura, Santiago, Chile}
\affiliation{Joint ALMA Observatory, Alonso de C\'ordova 3107, Vitacura, Santiago, Chile}

\email{paulo.cortes@alma.cl}

\begin{abstract}
We present the first linear-polarization mosaicked observations performed by the Atacama Large Millimeter/submillimeter Array (ALMA). We mapped the Orion-KLeinmann-Low (Orion-KL) nebula using super-sampled mosaics at 3.1 and 1.3\,mm as part of the ALMA Extension and Optimization of Capabilities (EOC) program. We derive the magnetic field morphology in the plane of the sky by assuming that dust grains are aligned with respect to the ambient magnetic field. At the center of the nebula, we find a quasi-radial magnetic field pattern that is aligned with the explosive CO outflow up to a radius of approximately $12^{\prime \prime}$ ($\sim$\,5000\,au), beyond which the pattern smoothly transitions into a quasi-hourglass shape resembling the morphology seen in larger-scale observations by the James-Clerk-Maxwell Telescope (JCMT). We estimate an average magnetic field strength $\left<B\right>=9.4$\,mG and a total magnetic energy of $2 \times 10^{45}$\,ergs, which is three orders of magnitude less than the energy in the explosive CO outflow. We conclude that the field has been overwhelmed by the outflow and that a shock is propagating from the center of the nebula, where the shock front is seen in the magnetic field lines at a distance of $\sim$\,5000\,au from the explosion center. \\

\end{abstract}

\section{INTRODUCTION}\label{se:intro}

The Orion nebula has been a subject of intense research since its discovery by Fabri de Peiresc in 1610 \citep{Miller2016}.
 Home to the Trapezium, a star cluster containing massive stars, the Orion nebula also hosts the closest high-mass star-forming  
region to the sun: the Orion integral-shaped filament \citep{Johnstone1999}. This one-degree-long, 2200\,\Msun\ filament is embedded in the 50,000\,\Msun\ Orion A cloud \citep{Bally1987}.  The $\sim$\,100\,\Msun\ Orion Molecular Core 1 (OMC1), located behind the Orion nebula \citep{Genzel1989} at a distance of $\sim$\,414\,pc \citep{Menten2007}, is the densest and most molecule-rich part of the Orion A cloud. 
OMC1 contains the Orion-KL nebula, which has been studied extensively over the past 40 years across the electromagnetic spectrum, and is composed 
of a number of different sources, of which the visually obscured Becklin–Neugebauer Object (BN), Source I, and source $x$ appear to be most dynamically relevant.
Note, that the BN, source I, and source $x$ are run-away stars moving at speeds between 10 to 30 {\kms} from a common origin
\citep{Goddi2011b,Luhman2017,Bally2020}. 

Orion-KL experienced an energetic explosion $\sim$\,500 years ago \citep{Gomez2008}\footnote{This enigmatic explosion has also been referred to as a ``magnetic bomb,'' coined in \citet{Bally2011}.}.
The origin of the explosion has been attributed to a gravitational interaction between three or more massive stars in the nebula, as suggested by the high-velocity proper motions of BN, source I, and source $x$ \citep{Bally2017,Bally2020}. This explosion affected the surrounding gas, as shown by vibrationally excited H$_{2}$ and [Fe {\small II}] emission that reveals a spectacular network of collimated outflows known as the ``Orion fingers'', which extend for 0.1\,pc in a wide-angle, conical configuration \citep{Allen1993, Colgan2007, Bally2015}. A plethora of observations of additional outflow tracers confirm this link between the outflow and the explosive event.  
In the millimeter, the fingers were detected in high-velocity CO emission by the Submillimeter Array (SMA)  \citep{Zapata2009}, and more recently with the Atcama Large Millimeter/Sub-millimeter Array (ALMA) \citep{Bally2017}
showing a spectacular, spherically symmetric, explosion profile. The center of the CO outflows is, indeed, located at the dynamical center of the massive stars thought to be involved in the explosion at the center of the Orion-KL nebula.

Because molecular clouds comprise weakly ionized plasma permeated by magnetic fields, it has long been an open question as to what is the interplay between the explosion and the magnetic field in OMC1. Initial studies with single dish telescopes captured the large-scale magnetic field structure in Orion-KL, revealing the well-known ``hourglass'' morphology, a cornerstone of magnetically controlled star formation  \citep{Schleuning1998,Ward-Thomson2017,Pattle2017,Chuss2019}.
To date, only interferometric polarization observations with limited sensitivity performed with the Berkeley-Illinois-Maryland Association (BIMA) array, the SMA, and the Combined Array for Research in Millimeter-wave Astronomy (CARMA) have been used to study the detailed magnetic field morphology of the Orion-KL system \citep{Rao1998,Tang2010,Hull2014}. Here, we expand on their initial analysis using the first\footnote{At time of submission, other ALMA observations from Orion-KL are reported, but un-refereed \citep[see ][]{Pattle2020}}  linear-polarization ALMA mosaic of Orion-KL, observed at high resolution and with high sensitivity. The paper is organized as follows: Section \ref{se:obs} describes the observations, calibration, and imaging; Section \ref{se:res} presents the results; Section \ref{se:discussion} presents the discussion; and Section \ref{se:conc} contains the conclusions and summary.

\begin{deluxetable*}{c c c c  c c c c c}[hbt!]
\tablecolumns{8}
\tablewidth{0pt}
\tablecaption{ Spectral Setup\label{table:setup}}
\tablehead{
\colhead{Band} &
\colhead{Spw 1} &
\colhead{Spw 2} &
\colhead{LO 1} &
\colhead{Spw 3} &
\colhead{Spw 4} &
\colhead{$B_\textrm{maj}$} &
\colhead{$B_\textrm{min}$} &
\colhead{$PA$} 
\\
\colhead{}     &
\colhead{[GHz]}     &
\colhead{[GHz]}     &
\colhead{[GHz]}     &
\colhead{[GHz]}     &
\colhead{[GHz]}     &
\colhead{[$''$]}  & 
\colhead{[$''$]}  &
\colhead{[$\arcdeg$]} 
}
\startdata
3 & 90.5 & 92.5 & 97.5  & 102.5 & 104.5 & 1.14 & 0.74 & $-59$ \\
6 & 224.0 & 226.0 & 233.0  & 240.0 & 242.0 & 0.74 & 0.53 & $-52$ \\
\enddata
\vspace{0.1cm}
\tablecomments{The spectral window (Spw) center frequency is indicated in GHz, along with the first local oscillator frequency (LO 1). The spectral resolutions of the Band 3 (3.1\,mm) and Band 6 (1.3\,mm) data provide 64 channels with 31\,MHz channel spacing. Additionally, we indicate the parameters of the synthesized beams (spatial resolution) of both observations: major axis $B_\textrm{maj}$, minor axis $B_\textrm{min}$, and position angle $PA$.}
\end{deluxetable*}

\section{OBSERVATIONS}\label{se:obs}

All of the data we present here and in the companion paper \citet{Hull2020b}
were collected as part of the ALMA Extension and Optimization of Capabilities (EOC) program.  
The EOC program is based at the Joint ALMA Observatory (JAO) in Santiago, Chile---with collaboration from scientists both at the ALMA Regional Centers (ARCs) as well as at external institutions---and is focused on using science-quality data to test, verify, and open new observing modes at ALMA.
The primary goal of these EOC observations was to assess the effects of off-axis instrumental polarization errors in maps of an extended, highly polarized source of scientific interest. We present a full description and characterization of the residual off-axis errors of the ALMA 12\,m antennas in \citet{Hull2020b}. 

As part of this EOC effort, we mapped the Orion-KL region with the 12\,m array in both bands 3 (3.1\,mm) and 6 (1.3\,mm) using the standard continuum setup for polarization observations (see Table \ref{table:setup} for a description).  The observations were configured as a rectangular, super-sampled mosaic with a factor of 2 over-sampling (with respect to Nyquist) along both axes, or a factor of 4 in total beam overlap.
Both mosaics have the same reference coordinates,
$\alpha$(J2000)\,=\,05:35:14.5, $\delta$(J2000)\,=\,$-$05:22:31.6. 
The observations at 3.1\,mm were performed between 17 and 18 March 2018 under marginal weather conditions.  The 1.3\,mm observations were performed on 12 April 2019 under regular weather conditions.  We calibrated the data using the standard procedure for ALMA polarization observations \citep[see][for a description]{Cortes2016,Nagai2016} using CASA version 5.4 \citep{McMullin2007}.
We imaged both mosaics by performing phase-only self-calibration with a final solution interval of 90\,s. All images were primary-beam corrected, and the polarized intensity image was debiased pixel-by-pixel \citep{Wardle1974, Hull2015}.
The final maps were produced using the Python-based \textit{APLpy} plotting package \citep{Robitaille2012}.
The resulting angular resolutions of the images are $\sim$\,$0\farcs9$ at 3.1\,mm and $\sim$\,$0\farcs6$  at 1.3\,mm (see Table\,\ref{table:setup}).

\section{RESULTS}\label{se:res}

\subsection{The total intensity mosaic of Orion-KL}\label{sse:si}

\begin{figure*}[hbt!]
\includegraphics[width=0.95\hsize]{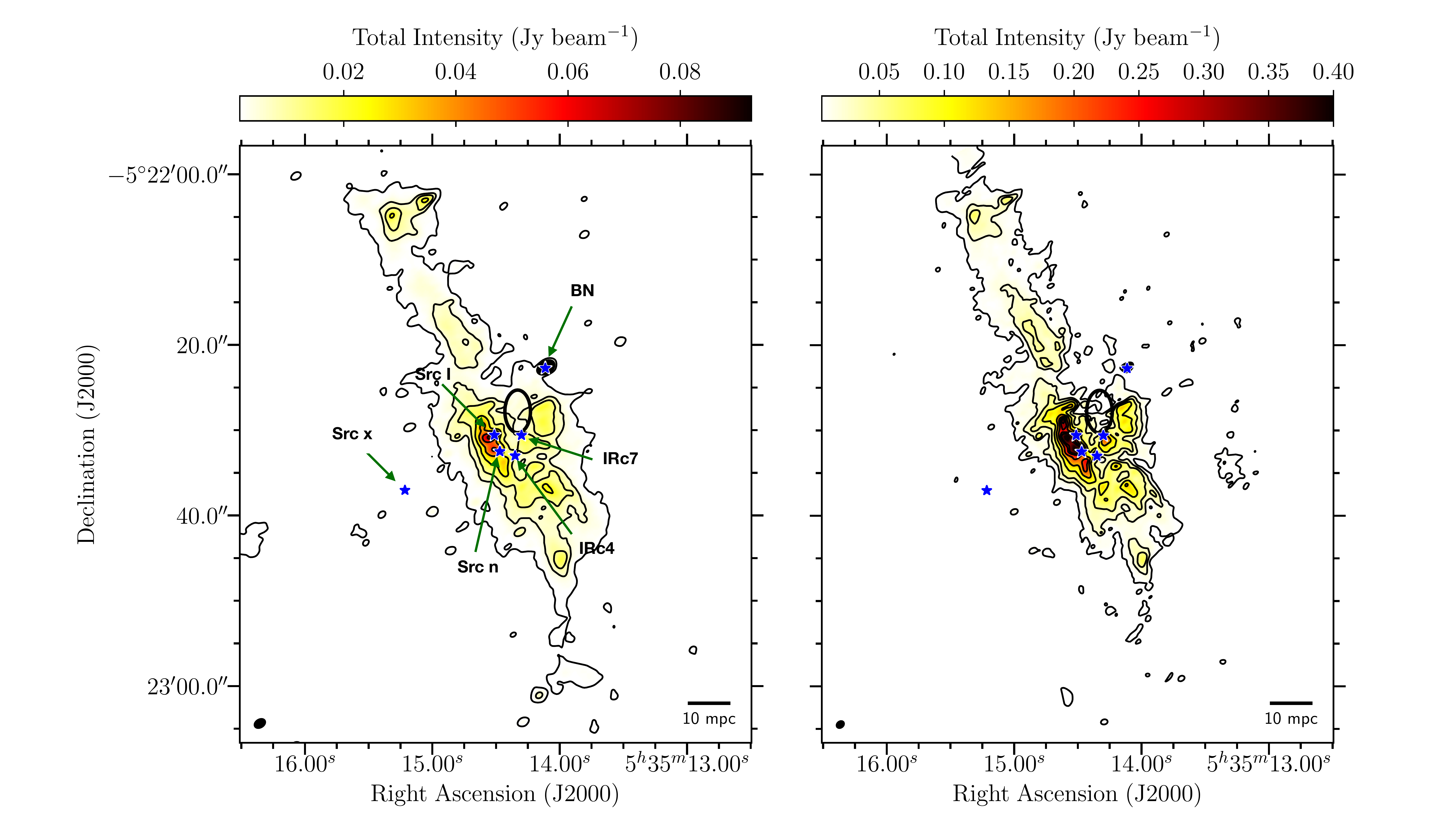}
\caption{Total intensity (Stokes $I$) emission maps toward Orion-KL. 
\textit{Left:} the 3.1\,mm data are in color scale with overlaid contours of 1.1, 5.5, 10.9, 21.8, 36.4, 54.6, 72.8, and 83.7\,mJy\,beam$^{-1}$. 
\textit{Right:} the 1.3\,mm data are in color scale with overlaid contours of 4.2, 21, 42, 84, 140, 210, 280, and 322\,mJy\,beam$^{-1}$.
Superposed on both maps are the most relevant sources, shown as blue stars and labeled in the 3.1\,mm map; their positions are from \citet{Bally2017,Bally2020}.
The black ellipse near the center of the image shows the proposed center for the explosion in Orion-KL \citep{Bally2017}.  
\label{fig:stokesI}} 
\end{figure*}

The maps of total intensity (Stokes $I$) emission from the 3.1\,mm and 1.3\,mm data are shown in Figure \ref{fig:stokesI}; these maps comprise the first-ever high-resolution ($\lesssim$\,1$''$) observations of the entire Orion-KL nebula. The mosaics cover regions of $\sim$\,$2.20^{\prime}\,\times\,1.75^{\prime}$ (15 pointings) at 3.1\,mm and $\sim$\,$1.25^{\prime}\,\times\,0.94^{\prime}$ (45 pointings) at 1.3\,mm, and are centered on Orion-KL.
The ALMA observations resolve the center of the nebula at both 3.1 and 1.3\,mm, where BN, source I, source $n$, and source $x$ are clearly visible, as indicated in the total intensity maps, with BN appearing as an isolated and unresolved object in both maps. We recover all of the continuum sources previously detected by others \citep{Eisner2006,Friedel2011,Hirota2015}. 

The morphology of the dust emission appears filamentary, consistent with previous ALMA 
observations of Orion-KL \citep{Hirota2015}. Toward the center of the nebula, we detect a 
cavity from the center of the CO outflow to the North in both the 3.1 and 1.3\,mm observations.  This cavity has a ``U'' shape and surrounds the center of the explosion; it is likely that it was created by the explosive event. 

We estimate the column density independently from the 3.1 and 1.3\,mm data by assuming optically thin dust emission:

\begin{equation}
        N_{\mathrm{H}_{2}} = \frac{S_{\nu}}{B_{\nu}(T_\textrm{dust}) \Omega \kappa_{\nu} \mu m_{\mathrm{H}}} \,\,, \\
\end{equation}

\noindent 
where $S_{\nu}$ is the total flux density of the source, $B_{\nu}(T)$ is the Planck function and $T_\textrm{dust}$ is the dust temperature. The area is given by the solid angle, calculated as $\Omega=\pi/(4.0\log{2})\,b_\textrm{maj}\,b_\textrm{min}$. The dust opacity $\kappa_{\nu}$ is set as 0.01\,cm$^{2}$\,g$^{-1}$
at 1.3\,mm and $2.6\times 10^{-3}$\,cm$^{2}$\,g$^{-1}$ at 3\,mm \citep{Ossenkopf1994}. These values assume a gas-to-dust ratio of 100. We use $\mu = 2.33$ as the mean molecular weight of hydrogen; $m_{\mathrm{H}}$ is the mass of a hydrogen atom. The optically thin assumption is justified, as preliminary spectral index estimations from these data suggest a spectral index of $\approx$\,3 in the region where we estimate the column density. For this work, we calculate the column density only in the region of the filament relevant to the estimation of the magnetic field strength (see Section \ref{sse:exp} and Table \ref{table:B}). We use the dust temperature maps from the JCMT reported by \citet{Salji2015a}. Based on that work, we use a range for $T_\textrm{dust}$ of between 20 and 50\,K. By using all the pixels in the two maps shown in Figure \ref{fig:stokesI}, we obtained total masses of 35 and
88.2 {\Msun} from the 3 and 1 mm data\footnote{We obtained cutoffs for the 3 and 1 mm Stokes $I$ maps of 900 $\mu$Jy beam$^{-1}$ and 7.3 mJy beam$^{-1}$ respectively.}, respectively by using $T_\textrm{dust} = 25$ K as an average dust temperature.

\begin{deluxetable*}{cccccccccccc}[hbt!]
\tablecolumns{12}
\tablewidth{0pt}
\tablecaption{Magnetic field estimation
\label{table:B}}
\tablehead{
\colhead{$\lambda$} &
\colhead{T$_{\mathrm{dust}}$} &
\colhead{$N$} &
\colhead{$n_{\mathrm{H_{2}}}$} &
\colhead{$\delta \phi$} &
\colhead{$\Delta V$} &
\colhead{$B_{1}$} &
\colhead{$B_{2}$} &
\colhead{$B_{3}$} &
\colhead{$\left< B_{\mathrm{pos}} \right>$} &
\colhead{$\sigma_{\mathrm{B}}$} & 
\colhead{$E_B$} \\
\colhead{[mm]} &
\colhead{[K]} &
\colhead{[$10^{24}$ cm$^{-2}$]} &
\colhead{[$10^{7}$ cm$^{-3}$]} &
\colhead{[degrees]} &
\colhead{[\kms]} &
\colhead{[mG]} &
\colhead{[mG]} &
\colhead{[mG]} &
\colhead{[mG]} &
\colhead{[mG]} &
\colhead{[$10^{45}$ erg]}
}
\startdata
3.1 & 20 & 3.0 & 24.5 &  19.7 & 1.4 & 10.3 & 16.8 & 9.5 & 12.2 & 3.3 &  3.3\\
3.1 & 30 & 1.8 & 14.8 &  19.7 & 1.4 & \phantom{0}8.0 & 13.1 & 7.4 & \phantom{0}9.5 & 2.6 &  2.0\\
3.1 & 40 & 1.3 & 10.6 &  19.7 & 1.4 & \phantom{0}6.8 & 11.1 & 6.2 & \phantom{0}8.0 & 2.2 &  1.4\\ \smallskip
3.1 & 50 & 1.0 & \phantom{0}8.3 &  19.7 & 1.4 & \phantom{0}6.0 & \phantom{0}9.8 & 5.5 & \phantom{0}7.1 & 1.9 &  1.1\\ 
1.3 & 20 & 5.4 & 44.0 &  21.1 & 1.4 & 12.9 & 20.9 & 2.6 & 12.1 & 7.5 &  3.2\\
1.3 & 30 & 3.5 & 28.2 &  21.1 & 1.4 & 10.3 & 16.7 & 2.1 & \phantom{0}9.7 & 6.0 &  2.1\\
1.3 & 40 & 2.6 & 20.7 &  21.1 & 1.4 & \phantom{0}8.9 & 14.3 & 1.8 & \phantom{0}8.3 & 5.1 &  1.5\\
1.3 & 50 & 2.0 & 16.4 &  21.1 & 1.4 & \phantom{0}7.9 & 12.8 & 1.6 & \phantom{0}7.4 & 4.6 &  1.2\\
\enddata
\vspace{0.1cm}
\tablecomments{Here we present the parameters used to estimate the magnetic field strength in Orion-KL, and the results of those estimates. 
We also show the value for the and energy ($E_B$) inside the 12$^{\prime \prime}$ radius where the magnetic field has a radial pattern.  The magnetic field energy was calculated using the average of the field strength onto the plane of the sky ($\left< B_{\mathrm{pos}} \right>$) as shown in column 10.  
}
\end{deluxetable*}

\begin{figure*} 
\includegraphics[width=0.95\hsize]{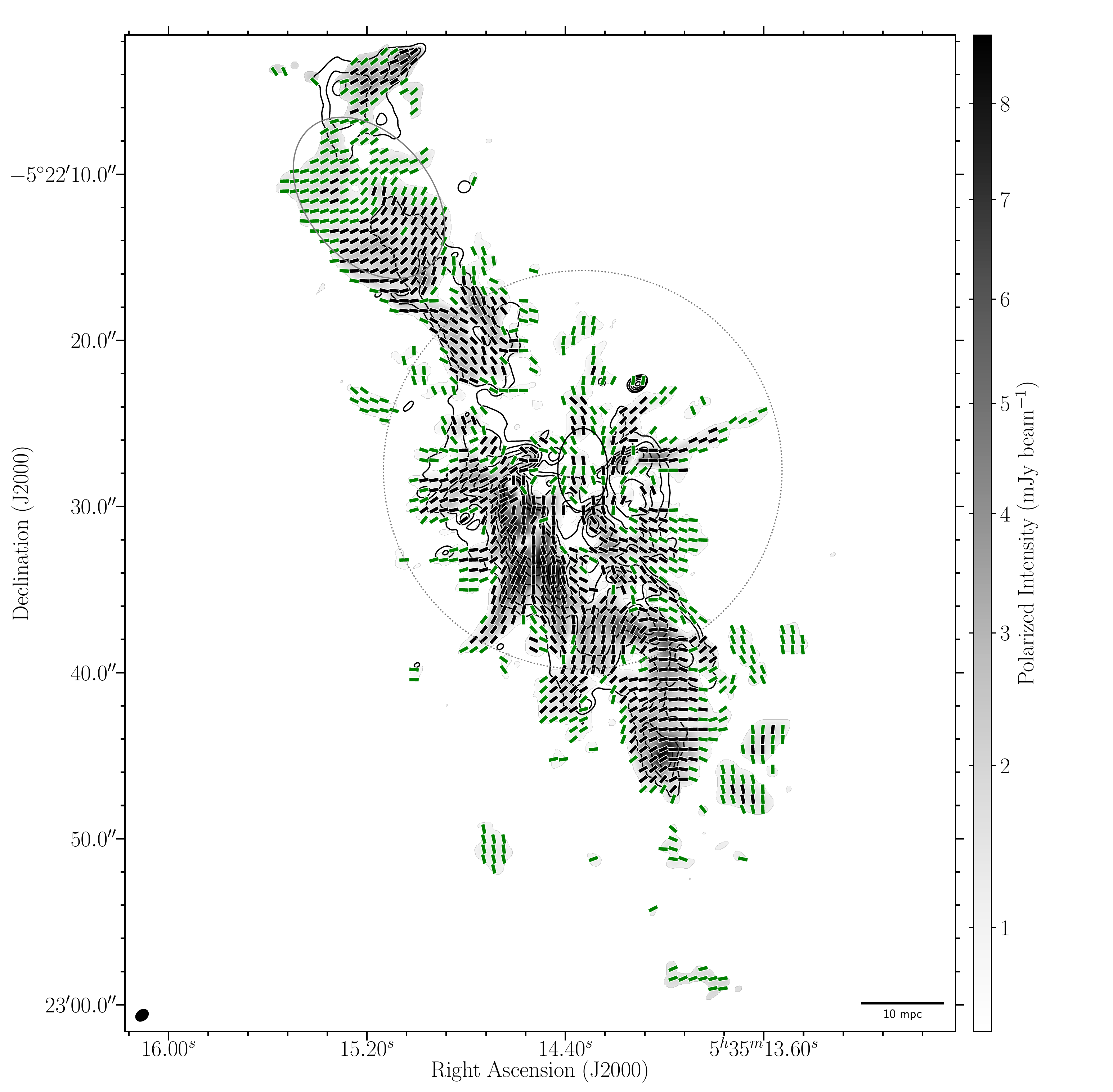}
\smallskip
\caption{Inferred magnetic field morphology in Orion-KL derived from the 1.3\,mm polarized dust emission. The green and black line segments correspond to polarization detections at the level of 3--5\,$\times$ and $>$\,5\,$\times$\,$\sigma = 200\,\mu$Jy, respectively. The line segments are normalized and plotted once per synthesized beam (coarser than Nyquist sampling by a factor of two in each dimension), where the beam size is $0.74^{\prime \prime}\,\times\,0.52^{\prime \prime}$; the beam is shown in the bottom-left corner as a solid, black ellipse. Gray scale is the polarized intensity in mJy, as indicated by the color bar, and is plotted from 600\,$\mu$Jy beam$^{-1}$. The contours correspond to the total intensity (Stokes $I$) emission at levels of 2.7,  13.5,  27,  54,  90, 135, 180, 207\,mJy\,beam$^{-1}$.  The black ellipse near the center indicates the $3^{\prime \prime}\,\times\,5^{\prime \prime}$ region proposed by \citet{Bally2017} to be the center of the explosion in Orion-KL. The gray dotted circle corresponds to the $\sim$\,12$\arcsec$ region where the magnetic field is perturbed and appears to have a quasi-radial profile. Finally, the gray ellipse shows the region where we estimate the magnetic field strength.
\label{fig:band6}
}
\bigskip
\end{figure*}

\subsection{The polarized dust emission from Orion-KL}\label{sse:spol}

Figure \ref{fig:band6} shows the 1.3\,mm mosaic of the magnetic field toward Orion-KL (the 3.1\,mm map is shown in Appendix \ref{se:app1}); these figures show the magnetic field morphology inferred from the polarized dust emission under the assumption that dust grains are magnetically aligned by radiative torques \citep[hereafter ``RATs,''; see][]{Lazarian2007}. We discuss the validity of this assumption in Section \ref{sse:galign}.
The magnetic field pattern is consistent between both wavelengths.  At 1.3\,mm, the polarized emission covers most of the filament sampled by the dust total intensity, where the derive magnetic field morphology is in agreement with  previous observations done at coarser angular resolution and/or lower sensitivity \citep[][see \citealt{Hull2020b} for more details]{Rao1998, Tang2010, Hull2014}.

\begin{figure*} 
\includegraphics[width=0.99\hsize]{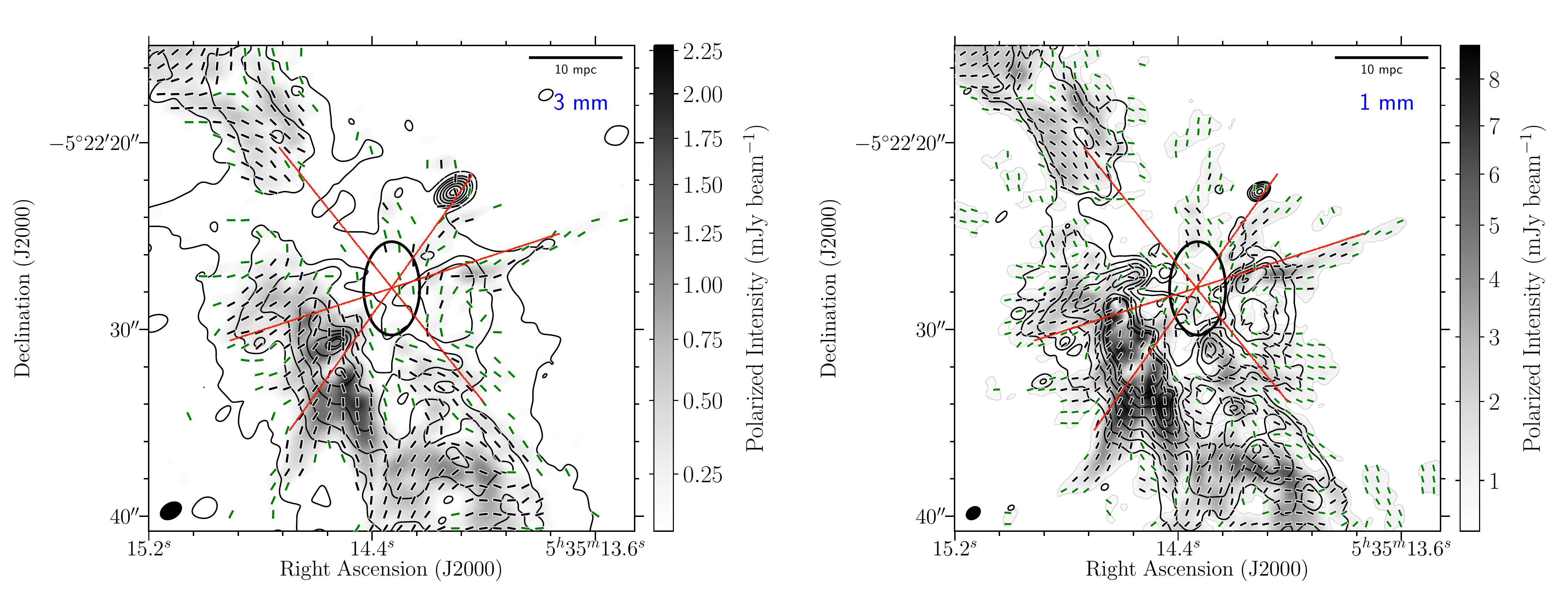}
\smallskip
\caption{A zoomed-in view of the magnetic field morphology in Orion-KL at 3.1\,mm (\textit{left}) and 1.3\,mm (\textit{right}).
Contours are the total-intensity Stokes $I$ dust emission, as shown in Figures \ref{fig:band6} and \ref{fig:band3}.
Over both maps, we have superposed the proposed center of the explosion as a black ellipse.
In orange lines, we superpose the extrapolation of the plane-of-sky field orientation to emphasize the quasi-radial morphology of the magnetic field in the center of the Orion-KL nebula.  
\label{fig:bombBfield}}
\bigskip
\end{figure*}

The magnetic field pattern can be divided into two distinct morphologies: (1) a component that appears quasi-perpendicular to the major axis of the filament, visible most clearly toward the northern extreme; and (2) a component at the center of Orion-KL, where the magnetic field morphology exhibits a pseudo-radial pattern emanating from the center of the nebula up to a distance of $\sim 12^{\prime \prime}$, or 5000 au (see Figure \ref{fig:band6}). By extrapolating the field lines through the center of Orion-KL, we see that they intersect inside the $3^{\prime \prime}\,\times\,5^{\prime \prime}$ ellipse reported to be origin of the explosive CO outflow \citep{Bally2017}. The extrapolation of the field lines is shown in Figure \ref{fig:bombBfield}. The transition from the quasi-radial morphology into an orientation quasi-perpendicular to the filament is smooth in both the northern and southern parts of the filament. Moreover, the quasi-perpendicular morphology resembles the hourglass shape seen by JCMT and the Stratospheric Observatory for Infrared Astronomy (SOFIA) at larger spatial scales \citep{Pattle2017,Chuss2019}. In fact, both morphologies, including the transition from radial to hourglass that we see in the ALMA data, can be seen in the SOFIA data \citep[see the 53\,$\mu$m map shown in the upper-left panel of Figure 1 in][]{Chuss2019}.

\subsection{The CO outflow and the magnetic field orientation}\label{sse:coB}

\begin{figure*} 
\includegraphics[width=0.99\hsize]{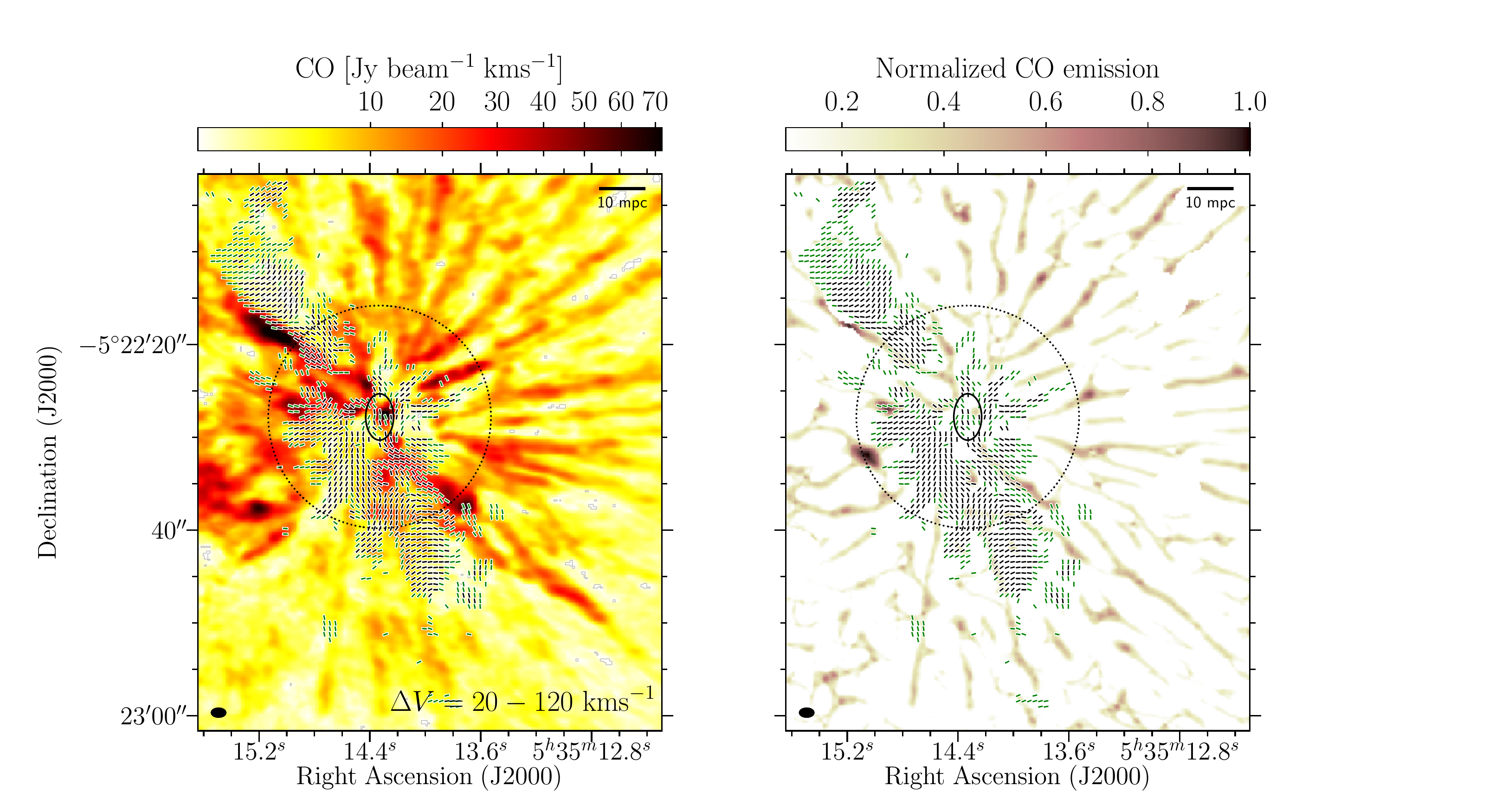}\\
\includegraphics[width=0.99\hsize]{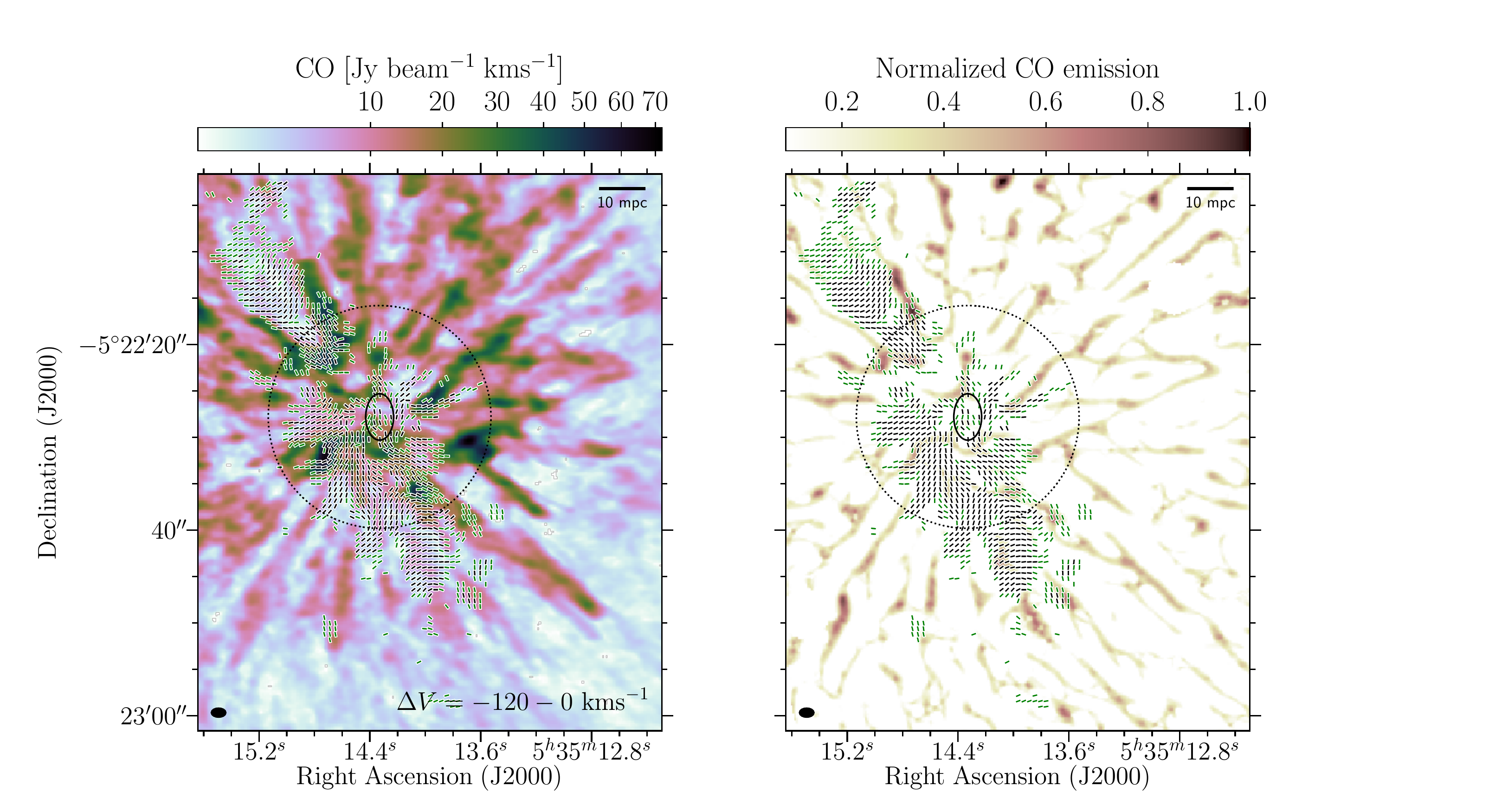}
\smallskip
\caption{Comparisons of the CO outflow, the magnetic field, and the RHT derived from the CO outflow (see section \ref{sse:coB}). \textit{Upper-left:} redshifted moment 0 map for the CO outflow, integrated over a range $\Delta V$ = 20 -- 120\,\kms{}, which removes the systemic-velocity emission from Orion-KL. \textit{Upper-right:} redshifted RHT map. \textit{Lower-left:} blueshifted emission moment 0 map of the CO outflow, integrated over $\Delta V$ = --120 -- 0\,\kms{}. \textit{Lower-right:} blueshifted RHT map.  As in previous maps, the magnetic field morphology is superposed on all four panels (see Figure \ref{fig:band6} for a description).
}
\label{fig:bomb6}
\bigskip
\end{figure*}

Figure \ref{fig:bomb6} shows the comparison between the magnetic field morphology and the spherically symmetric CO outflow. The magnetic field is superposed over the $^{12}$CO$\,(J=2\rightarrow1)$ moment 0 maps of both the red- and blueshifted outflowing CO emission \citep[see][]{Bally2017}. 
In many locations, the magnetic field seems to be aligned with the radial CO streamers, particularly to the West, where the field exhibits prominent radial, ``finger''-like structures.  Although the radial orientation seems clear, the locations where we detect dust polarization seem to be anti-correlated with the locations of the CO streamers. This anti-correlation is seen in maps of both the red- and blueshifted CO emission (see Figure \ref{fig:bomb6}). 
Outflows carving cavities in the dust have been seen in other star forming regions; in several of these, the magnetic field has been seen to be elongated along the cavities of the outflow, similar to what we see in Figure \ref{fig:bomb6} \citep{Hull2017b,Maury2018,Takahashi2019,LeGouellec2019,Hull2020a}.

Comparing with the dust emission, we see that the finger-like structures in the CO extend to scales $\sim$\,3\,$\times$ larger than the region where we detect the total intensity dust emission (this can be seen by comparing Figures \ref{fig:stokesI} and \ref{fig:bomb6}), in particular in directions perpendicular to the filament's major axis. Toward the center of the explosion, the field morphology is more complicated: the field is obviously perturbed, with clear departures from
the large-scale field morphology. 

To further explore the alignment between the CO streamers and the magnetic field, we use the Rolling Hough Transform \citep[or RHT, see][]{Clark2014}. The RHT is a machine-vision algorithm that is well suited for the identification of linear structures in data (e.g., edge detection). We applied this algorithm to the CO streamers to highlight the linear structures in the data and to correlate them with our polarization map. The linear structures in the CO streamers are clearly identified by the RHT algorithm in both the red- and blue-shifted lobes of the outflow (see the right-hand panels in Figure \ref{fig:bomb6}). Comparing these structures with the magnetic field, we can clearly see a correlation in the orientations, in particular as we move radially away from the center of the explosion. The RHT maps also confirm the anti-correlation between the dust and CO emission that is seen to the West of the Orion-KL nebula (and also to the East, to some degree), as previously mentioned.  
In addition, we clearly see radially oriented CO emission in regions where the magnetic field appears to be unperturbed: i.e., toward the ends of the filament where the dust morphology is compact, filamentary, and threaded by a quasi-hourglass field shape.

\subsection{Estimation of the magnetic field strength}\label{sse:exp}

To estimate the strength of the magnetic field, we use three versions of the Davis-Chandrasekhar-Fermi method (DCF; \citealt{Davis1951,Chandrasekhar1953}), as previously employed in \citet{Cortes2016,Cortes2019}. The DCF technique assumes that the magnetic field lines are perturbed by non-thermal turbulent motions, and thus we cannot directly estimate the field strength from the central region of Orion-KL that has been affected by the explosion. This is because the explosion injects an amount of energy such that the field lines perturbations are likely   non-linear,
which makes the assumptions in the DCF technique not applicable. Instead, we use the NE region of the filament, where the field appears to be unperturbed by the explosion (indicated by the gray ellipse in  Figure \ref{fig:band6}). The DCF method requires estimates of the gas number density, the velocity dispersion $\Delta V$, and the dispersion in the magnetic field lines, or $\delta \phi$. To estimate the gas density, we assume a thin slab with a depth of half the region's minor axis, or $n = N/L$ with $L = 2^{\prime \prime}$,
where $N$ is the estimated column density (see Table~\ref{table:B}).  For the velocity dispersion, we use the NH$_{3}$ observations from \citet{Wiseman1998} performed using the Karl G. Jansky Very Large Array (VLA) at 8$^{\prime \prime}$ resolution. They obtained a value of 1.4\,\kms{} at the location of Orion-KL (see their Table 2).  Given that their resolution approximately matches the minor axis of our region of interest, we used that value for the line-width in our field-strength estimation.  The value of $\delta \phi$ is calculated from our polarization data directly. Because the 3.1 and 1.3\,mm data trace different length scales, we derive independent estimations of the field strength from each dataset, and obtain average strengths between 7 and 12\,mG. By averaging all of the estimates from Column 10 in Table \ref{table:B}, we obtain an average strength at 1.3\,mm value of $\left<B\right>=9.4$\,mG. Note that these three flavors of the DCF method do not take into account the effect of finite angular resolution and integration along the line of sight, both of which tend to smooth the magnetic field and reduce angle dispersions, and thus we
consider this estimate as an upper limit of the magnetic field strength. We present the results of these estimations and the parameters used in Table \ref{table:B}.

\section{DISCUSSION}\label{se:discussion}

\subsection{Grain alignment}\label{sse:galign}

One of our main  assumptions is that grains are aligned by magnetic fields. The other plausible mechanism---mechanical alignment---has two flavors. The classical mechanism requires a significant dust-gas drift \citep{Gold1952}, and appears to be important only for highly irregular dust grains. \citet{Tang2010} discarded the classical Gold mechanism, arguing that the polarization pattern should be aligned with the flow; this is the opposite of the azimuthal polarization pattern that we see in Orion-KL.\footnote{Here we refer to the polarization position angle pattern, not to the magnetic field pattern; we obtain the latter by rotating the former by 90$^{\circ}$.}
The second flavor was introduced by \citet{Hoang2018} and is called ``Mechanical Alignment Torques'' (or MATs), which are proposed to operate in high-mass regions where radiation pressure drives the outflow. 
The spatial anti-correlation between the outflow fingers and the regions of radial magnetic field (as traced by the polarized dust emission) suggests that this mechanism is not occurring, as the CO and dust emission would need to be co-spatial for mechanical alignment to be plausible.  Therefore, it seems that the most likely mechanism to explain the polarized emission is magnetically aligned dust grains.

\subsection{Energetics comparison: outflow, gravity, and magnetic field}

The quasi-radial morphology of the magnetic field in Orion-KL suggests significant alignment of the magnetic field (projected onto the plane of the sky) with the outflow. A question that immediately emerges is whether this is because the field lines have been bent by the gas as it moved outward, or whether the field lines have been dragged inward by gravity. Radial magnetic field configurations have been attributed to gravitational collapse in other high-mass star-forming regions such as W51 \citep{Tang2009,Koch2018} and W43-Main \citep{Cortes2016,Cortes2019,ArceTord2020}. We will consider this question by evaluating the energy balance in the region surrounding the Orion-KL nebula. 

\citet{Bally2020} estimated a total mass for the Orion-KL complex of $\sim$ 100\,M$_{\odot}$, which yields
a gravitational potential energy of E$_{W} = GM^{2}/R = 3.52 \times 10^{46}$ ergs, when considering a sphere of  $12^{\prime \prime}$ (5000 au) in radius enclosing the region where the field appears perturbed. This energy estimate is 2 orders of magnitude less that the maximum energy estimated in the outflow, or $\sim 10^{48}$\,erg \citep{Bally2017}. 
The magnetic field tension must be overcome in order to explain the quasi-radial morphology that we see in our maps.  To test whether this is occurring, we must estimate the total magnetic energy in the perturbed region E$_{\mathrm{B}}$ = $B^{2}R^{3}/12\pi \sim 2 \times 10^{45}$\,erg, where we used a magnetic field strength of 9.4\,mG. This energy value is $\sim$\,3 orders of magnitude less than the maximum estimated energy of the outflow. Thus, the energy in the explosion is sufficient to overcome even a magnetic field with a relatively large ($\sim$\,10\,mG) strength. 

From this simple energetics comparison, it appears that the outflow is the dominant dynamical player in the region immediately surrounding the explosion center.
However, it is important to remember that the magnetic field ``feels'' dynamical interactions through collisions with the charge carriers, as it is likely that the field is ``frozen'' into the charged fluid \citep{Kulsrud2005}. Both fluids, neutral and charged, are subject to the gravitational field, but for the magnetic field to be dynamically perturbed by the outflow
we need to estimate how well the charge carriers are coupled to the neutrals. To quantify this, we estimate the magnetic Reynolds number $R_{m}$. Values of $R_{m} \gg 1$ suggest strong coupling between the gas and the magnetic field; under such a scenario, the gas and the magnetic field should move as a single fluid. 
We estimate a value of  $R_{m} = 1.7 \times 10^{5} \gg 1$ by considering the same quantities used for the estimation of the magnetic field strength (see
Appendix \ref{se:mren} for details).  Similar values for $R_{m}$ have been estimated in the literature towards giant molecular clouds \citep[or GMCs,][]{Myers1995}. Even if the magnetic field strength were a factor of 10 less ($\sim$\,1\,mG), we would still obtain $R_{m} \gg 1$. Based on this result, we conclude that the field is well coupled with the gas. Therefore, considering the difference in magnitude between the energies in the outflow and in the gravitational field, it is more likely that magnetic field morphology is the result of the field being stretched outward by the outflow rather than being pulled inward by gravity. This is particularly clear at the locations of the CO ``fingers,'' which are located away from the main filament where the gravitational pull should be less important.

\subsection{A Shock in the Orion-KL Magnetic Field}\label{sse:cshock}

Among the plethora of features seen in emission from Orion-KL, the vibrationally excited H$_{2}$ and [Fe {\small II}] emissions are the most visually striking \citep[][see also Figure \ref{fig:band3H2}]{Allen1993, Nissen2007, Bally2011, Bally2015}. Vibrationally excited H$_{2}$ emission is a signpost of shock activity because this emission is one of the main coolants for the post-shock gas \citep{Draine1993}, while [Fe {\small II}] emission is commonly seen toward Herbig-Haro objects. Additionally, observations of emission from complex organic molecules (or COMs) toward the Orion-KL region strongly suggest that shocks have affected the chemistry. Because a shock injects a significant amount of kinetic energy into the medium, there can be increased production of molecules that are not normally abundant in quiescent gas. This is supported by BIMA, SMA, IRAM 30\,m, and ALMA observations of a number of shock tracers such as SO$_{2}$, $^{34}$SO$_{2}$, SO, and heated H$_{2}$CS, and also by the discovery of chemical segregation in COMs towards Orion-KL \citep{Liu2002,Feng2015,Favre2017,Pagani2017,Tercero2018,Pagani2019,Li2020}. 
Recently, \citet{Pagani2019} found elongated formaldehyde (H$_{2}$CO) emission whose origin is consistent with that of the explosive event in Orion-KL. This is critical, because the elongation direction of the H$_{2}$CO correlates well with the radial morphology seen in the CO outflow and with the quasi-radial magnetic field morphology we are presenting here. We further note that the maximum elongation of the H$_{2}$CO emission (relative to the center of the explosive event) is approximately the same as the radius of the transition from quasi-radial to quasi-hourglass seen in the magnetic field morphology. 

\begin{figure*}
\includegraphics[width=0.95\hsize]{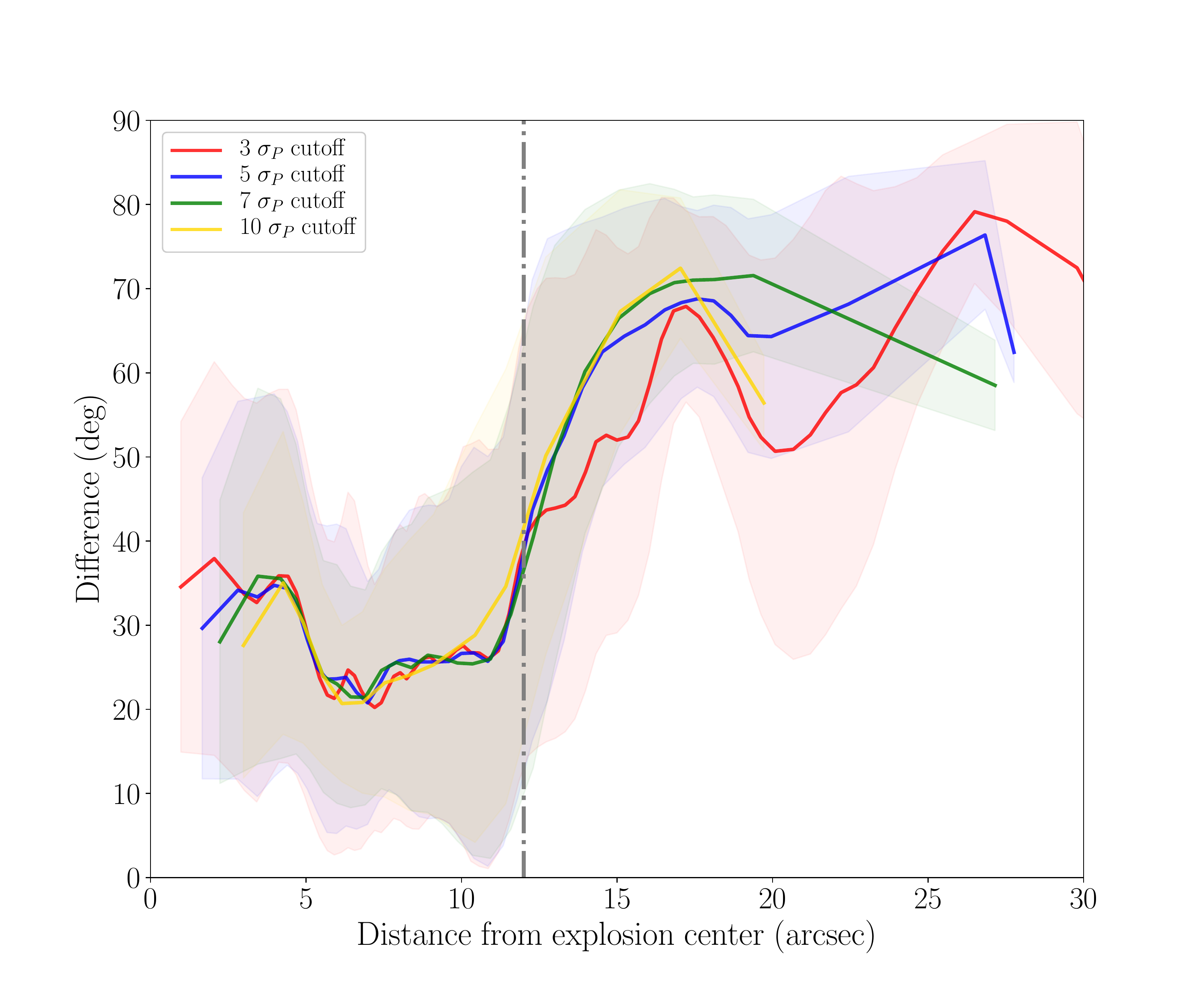}
\smallskip
\caption{The profile of the difference in positions angle between a purely radial magnetic field (i.e., a monopole field morphology centered at the explosion center reported in \citealt{Bally2017}) and the magnetic field morphology derived from the ALMA 1.3\,mm polarized dust emission (Figure \ref{fig:band6}). The maps have been regridded to a Nyquist pattern (4 pixels per beam area). The colors represent the significance in the cuts (where $\sigma_P$ is the noise of the polarized intensity map) taken when binning and averaging the position angle data.  The shades of the same color represent the error calculated as the standard deviation of each bin. The vertical dashed line indicates the position of the potential shock front.
\label{fig:profile}
} 
\bigskip
\end{figure*}

\begin{figure*}
\includegraphics[width=0.95\hsize]{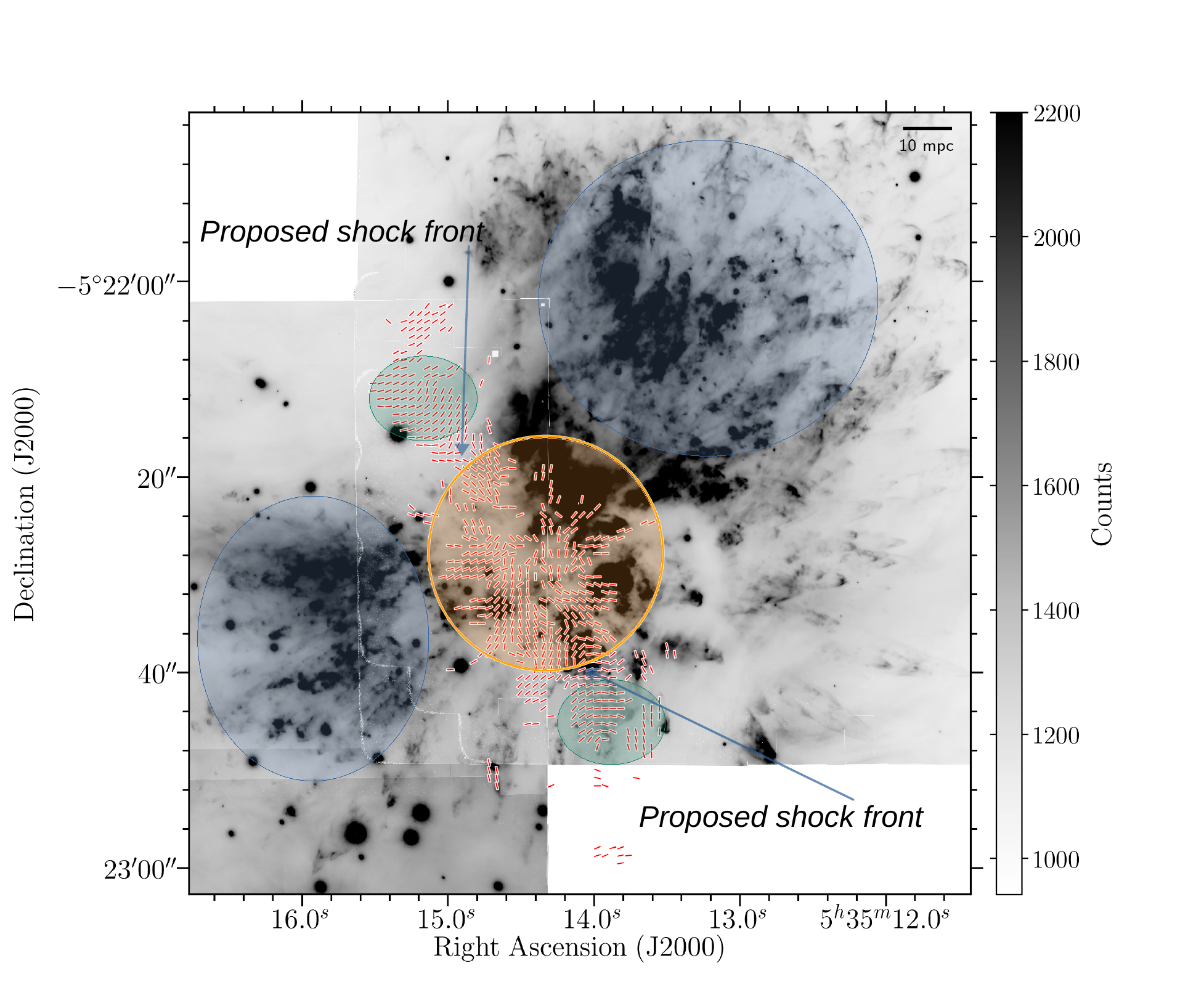}
\smallskip
\caption{H$_{2}$ ro-vibrational emission from \citet{Bally2015} with the superposed magnetic field morphology in red pseudo-vectors as derived from the 1\,mm data.  The orange semi-transparent circle marks the 12$^{\prime \prime}$ boundary for the region where the magnetic field appears perturbed. The blue semi-transparent circles show the peak of emission in H$_{2}$ while the the green  semi-transparent circles  show the unperturbed magnetic field regions. The arrows show the approximated proposed places for the shock front along the dust filament.
\label{fig:band3H2}
} 
\bigskip
\end{figure*}

As previously mentioned, the magnetic field exhibits a quasi-radial morphology only up to scales of 12$^{\prime \prime}$ ($\sim$\,5000\,au), whereas both the outflow and the dust emission in the filament extend well beyond that radius. 
At this $\sim$\,5000\,au radius, the field seems to continuously transition to the quasi-hourglass morphology seen on large scales by JCMT and SOFIA, far from the center of the explosion. We propose that this transition marks the position of a  shock front that is propagating from the origin of the explosion in Orion-KL. To visualize this transition better, we plot a profile for the magnetic field as a function of distance from the center of the explosion. We present this in Figure \ref{fig:profile}, where we show the difference between a purely monopole magnetic field (i.e., a radial profile) and the field position angle from the 1.3\,mm data. The quasi-radial morphology within the
proposed shocked region is clearly seen, having an almost constant average difference value of $\sim 25^{\circ}$; the transition at a radius of $\sim$\,12$^{\prime \prime}$ manifests itself as a sharp slope in the difference profile.
Furthermore, using the time and location of the explosive event as  initial conditions, we obtain a speed of $\sim 50$ {\kms } for a shock traveling along the dust
filament, up to 12$^{\prime \prime}$  in Orion-KL. Although the area in which the field appears perturbed (i.e., primarily radial) is clearly within a circle of $\sim$\,12$^{\prime \prime}$ in radius, the shock does not have to be spherically symmetric 
because the shock speed will depend on the ambient density in
which the perturbation is propagating. 
\citet{Bally2015} presented H$_{2}$ and [Fe {\small II}] emission that show ``fingers'' and wakes that are $\gtrsim 30^{\prime \prime}$ 
away from the explosion center and the dust filament. They 
suggests a perturbation traveling at high speeds (between 200 and 300 {\kms}), as measured from the proper
motions of the H$_{2}$ and [Fe {\small II}] ``bullets'' whose origin is also consistent with the origin of CO explosive outflow. 
We argue that this 
perturbation, or shock moving at $\sim 50$ {\kms}, is slower because it is traveling through the much
denser environment probed by the ALMA observations.

To further support our proposal, we compare our 1.3 mm magnetic field map with the aforementioned H$_{2}$ emission map from Orion-KL \citep{Bally2015}. Figure \ref{fig:band3H2} shows the magnetic field map superposed on the H$_{2}$ emission with semi-transparent ellipses indicating  the different regions affected by the shock. 
The brightest region of H$_{2}$ emission is located outside the dust filament, where dust emission levels are below the ALMA detection threshold. When we look at our proposed post-shock region (as indicated by the orange semi-transparent circle in Figure \ref{fig:band3H2}), we do see diffuse H$_{2}$ emission in this post-shock region, which is expected, as vibrationally excited H$_{2}$ emission is a major coolant in shocked regions.
However, beyond the shock front (i.e., the $\sim$\,12$^{\prime \prime}$ boundary), the H$_{2}$ emission appears significantly attenuated, in particular as we move NE along the main axis of the dust filament and away from the post-shock region into the regions of unperturbed magnetic field, which are shown as green semi-transparent ellipses in Figure \ref{fig:band3H2}.  

Unfortunately, from our data alone, we cannot conclude what type of shock is manifesting itself through the magnetic field morphology in Orion-KL.
Furthermore, to the best of our knowledge,  current models are only one-dimensional, with the field orthogonal to the shock propagation direction \citep[with the notable exception of][who computed models using oblique magnetic fields with respect to the shock front]{Pilipp1994}; therefore, we do not have a model to which we can refer for comparison. In particular, explaining the smooth transition seen in the magnetic field morphology and the field's relationship with the shock front will require such models. Nonetheless, we hope that these findings
will motivate much-needed multi-dimensional numerical work.

\section{SUMMARY AND CONCLUSIONS}\label{se:conc}

We present the first linear-polarization mosaic performed by ALMA, targeting polarized dust emission at 3.1 and 1.3\,mm in the Orion-KL star-forming region. From these data, we derive the plane-of-sky magnetic field morphology, which we compare with the explosive CO outflow emanating from the center of the nebula. We also estimate the strength of the magnetic field and evaluate the energy balance in the source. Our conclusions are as follows:

\begin{enumerate}

\item We see a quasi-radial magnetic field morphology in Orion-KL up to a radius of $\sim$\,12$^{\prime \prime}$
from the center of the explosive event. This radial pattern in the magnetic field is consistent with the radial orientation of the explosive CO outflow.  

\item The radial patterns in the magnetic field and CO outflow have the same origin; this location is spatially consistent with the origin of the dynamical event that triggered the explosion $\sim$\,500 years ago.

\item A comparison of the energy in the magnetic field and the outflow suggests that the outflow energy by far dominates that in the magnetic field, which can explain the quasi-radial orientation in the inner 12$^{\prime \prime}$.  

\item We propose that the quasi-radial morphology of the magnetic field is tracing a shock, where the shock-front itself appears to be very narrow. Here, the radial profile traces the post-shocked material, which is further supported by chemical studies showing shock-cooling molecular emission in this same region. 

\end{enumerate}

ALMA has revealed in exquisite detail the magnetic field in Orion-KL for the first time by mosaicking the polarized dust continuum emission. Further observations and modelling will be needed to improve our understanding of the enigmatic explosion in Orion-KL.

\acknowledgements
The authors are most grateful for the critical comments provided by Professor John Bally who refereed this paper.
The authors also thank A.J. Maury for the helpful comments and insights into this work.
P.C.C. acknowledges publication support from ALMA and NRAO.
V.J.M.L.G. and C.L.H.H. acknowledge the ESO Studentship Program.
C.L.H.H. acknowledges the support of the NAOJ Fellowship and JSPS KAKENHI grant 20K14527.
C.L.H.H. and J.M.G. acknowledge the support of JSPS KAKENHI grant 18K13586.
J.M.G. acknowledges the support of the grant AYA2017-84390-C2-R (AEI/FEDER, EU).
This paper makes use of the following ALMA data: ADS/JAO.ALMA\#0000.0.00377.CSV.
ALMA is a partnership of ESO (representing its member states), NSF (USA) and NINS (Japan), together with NRC (Canada), MOST and ASIAA (Taiwan), and KASI (Republic of Korea), in cooperation with the Republic of Chile. The Joint ALMA Observatory is operated by ESO, AUI/NRAO and NAOJ.
The National Radio Astronomy Observatory is a facility of the National Science Foundation operated under cooperative agreement by Associated Universities, Inc. 

\textit{Facilities:} ALMA.

\textit{Software:} APLpy, an open-source plotting package for Python hosted at \url{http://aplpy.github.com} \citep{Robitaille2012}.  CASA \citep{McMullin2007}.  Astropy \citep{Astropy2018}.

\bibliographystyle{apj}
\bibliography{biblio}

\clearpage

\appendix

\section{The 3.1\,mm mosaic}\label{se:app1}

\begin{figure*}[hbt!]
\includegraphics[width=0.95\hsize]{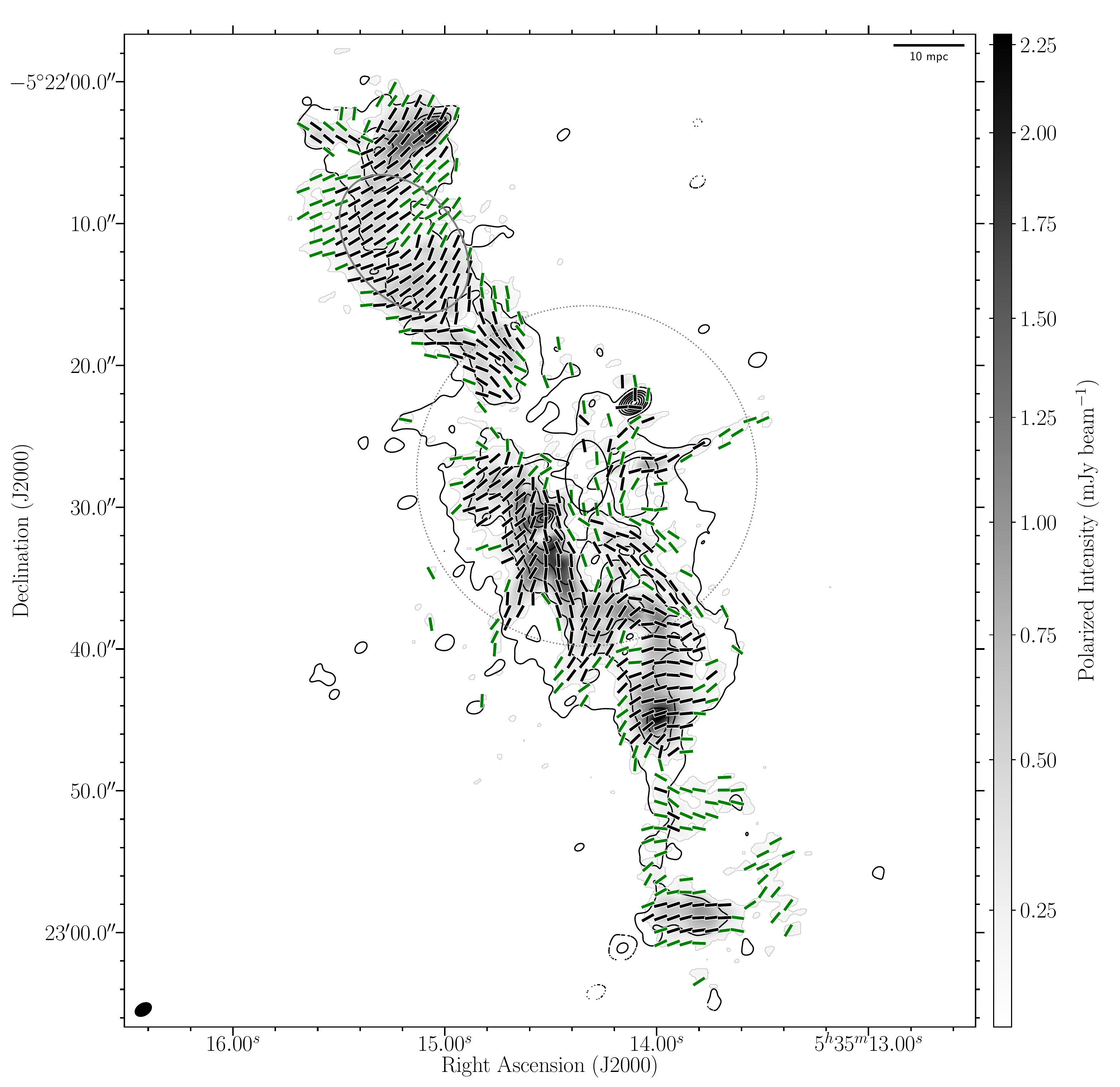}
\smallskip
\caption{Inferred magnetic field morphology toward Orion-KL derived from the 3.1\,mm dust emission, plotted in the same manner as Figure \ref{fig:band6}. Contours indicate the total intensity, with the same levels as Figure \ref{fig:stokesI}.  Gray scale corresponds to the polarized intensity, and is plotting starting at 126\,$\mu$Jy beam$^{-1}$. Line segments are the inferred magnetic field morphology.
The green and black line segments correspond to 3--5\,$\times$ and >\,5\,$\times$\,$\sigma$ = 42\,$\mu$Jy beam$^{-1}$.
\label{fig:band3}
} 
\bigskip
\end{figure*}

In Figure \ref{fig:band3} we present the 3.1\,mm map of the inferred magnetic field toward Orion-KL. As was the case for the 1.3\,mm observations (see Figure \ref{fig:band6}), the 3.1\,mm observations used a super-sampled mosaic pattern. A complete analysis of these data, in the scope of ALMA calibration, is presented in \citep{Hull2020b}. 

\clearpage

\section{The magnetic Reynolds Number}\label{se:mren}

The magnetic Reynolds number can be written as $R_{m} = v_{f}l/\nu_{M}$, where $v_{f}$ is the velocity of the fluid, $l$ is the length across which the magnetic field strength can change significantly, and $\nu_{M}$ is the magnetic viscosity, defined as $\nu_{M} = v_{A}^{2}/\nu_\textrm{in}$. Here, $v_{A}$ is the Alfv\'en speed and $\nu_\textrm{in}$ is the ion-neutral collision frequency, defined as $\nu_\textrm{in} = X(e)n_{\mathrm{H}}\left<\sigma_{\mathrm{in}} v\right>$. Here, $X(e)$ is the electron fraction and $\left< \sigma_{\textrm{in}} v \right> $ is the ion-neutral collision rate.  Thus, we can write the magnetic Reynolds number as 

\begin{equation}
    R_{m} = \frac{v_{f}l}{v_{A}^{2}} X(e)n_{\mathrm{H}} \left< \sigma_{\mathrm{in}} v \right>\,
\end{equation}

\noindent
To calculate $R_{m}$, we use the same values used to estimate the magnetic field strength (see section \ref{sse:exp}. Thus, $v_{f} = 1.4$ \kms{}, $l = 12^{\prime \prime}$ or 0.02\,pc, $v_{A} = 0.12$\,\kms{} considering a 10\,mG field, $n_{\mathrm{H_2}} = 3.0 \times 10^{8}$\,cm$^{-3}$ (see Table \ref{table:B}), $\left<\sigma_\textrm{in}v\right> = 1.69 \times 10^{-9}$\,cm$^{3}$\,s$^{-1}$ \citep{Mouschovias1981}, and $X(e) = \mathrm{N}_{e}/N_{\mathrm{H_2}}$, where N$_{e}$ = 7.42$ \times 10^{17}$\,cm$^{-2}$ is taken from estimations of rotation measures toward Orion A by \citet{Tahani2018}. This yields $X(e) \sim 10^{-7}$, which is most likely a lower limit, as the radiation field
produced by the nearby stars and enhanced by the explosion might have increased the electron density. Using these values we obtain $R_{m} = 1.7 \times 10^{5} \gg 1$, which means that there is strong coupling between the gas and the magnetic field; under such a scenario, the gas and the magnetic field should move as a single fluid.

\end{document}